\documentclass[fleqn,10pt]{wlscirep}
\usepackage[utf8]{inputenc}
\usepackage[T1]{fontenc}
\usepackage{hyperref}
\usepackage{algorithm}      % Provides the floating 'algorithm' environment
\usepackage{algorithmic} 

\title{Pair Approximation Meets Reality: Diffusion of Innovation in Organizational Networks within the biased-independence q-Voter Model}

\author[1,*]{Angelika Abramiuk-Szurlej}
\author[1,+]{Katarzyna Sznajd-Weron}

\affil[1]{Faculty of Management, Wroc\l{}aw University of Science and Technology, 50-370 Wroc\l{}aw, Poland}

\affil[*]{angelika.abramiuk-szurlej@pwr.edu.pl}
\affil[+]{katarzyna.weron@pwr.edu.pl}


\begin{abstract}
Collective adaptation, whether in innovation adoption, pro-environmental or organizational change, emerges from the interplay between individual decisions and social influence. Agent-based modeling provides a useful tool for studying such processes. Here, we introduce the biased-independence $q$-voter model, a generalization of the $q$-voter model with independence, one of the most popular agent-based models of opinion dynamics. In our model, individuals choose between two options, adopt or not adopt, under the competing influences of conformity and independent choice. Independent choice between two options is determined by an engagement parameter, inspired by earlier agent-based model of eco-innovation diffusion. When the engagement parameter equals $0.5$, the model reduces to the original $q$-voter model with independence; values different from $0.5$ break the symmetry between the two options. To place our study in a broader context, we briefly review asymmetric versions of the $q$-voter model proposed to date. The novelty of this work goes beyond introducing a generalized model: we develop the pair approximation (PA) for an asymmetric $q$-voter model and, for the first time, validate it on empirical organizational networks. Our results show that the interplay of social influence, independence, and option preference generates discontinuous phase transitions and irreversible hysteresis, reflecting path-dependent adoption dynamics. Surprisingly, the PA agrees well with Monte Carlo simulations on some empirical networks, even small ones, highlighting its potential as a computationally efficient bridge between individual decision-making and collective actions.
\end{abstract}
\begin{document}

\flushbottom
\maketitle

\thispagestyle{empty}

%%%%
\section{Introduction}
\label{sec:Introduction}
Diffusion of innovation is generally understood as the spread of new ideas, products, and practices over time among the members of a given social system. Numerous empirical examples demonstrate that even clearly beneficial innovations can spread slowly or fail altogether \cite{Rogers2003DiffusionEdition}. Similar patterns have been observed in collective adaptation, where path dependence (hysteresis), lack of optimization, and collective myopia can hinder the effective responses of collectives to new challenges \cite{Galesic2023BeyondAdaptation}. Therefore, in today’s rapidly changing world, understanding the factors that influence such forms of collective action is of great importance. In this paper, we focus on the diffusion of innovation as a concrete example of collective adaptation, while noting that our model is sufficiently general to capture any situation in which a population initially adheres to an old state and must collectively adjust to a new one. 

Over the years, various approaches have been used to study the diffusion of innovation, but since the beginning of this century, agent-based modeling (ABM) has gained increasing popularity \cite{Garcia2005UsesResearch,Rand2011Agent-BasedRigor,Karsai2014ComplexInnovation,Rand2021AgentbasedCriticisms,Mueller2023SignedInnovation,Stummer2024Agent-BasedManagement}. One of the ABMs used to describe the diffusion of eco-innovation \cite{Byrka2016DifficultyPractices} is based on the $q$-voter model \cite{Castellano2009NonlinearModel}. In the $q$-voter model, individuals (agents) hold binary opinions $\pm 1$, representing yes/no, agree/disagree, adoption/non-adoption, and so on. They update their opinions according to the influence of a group of $q$ neighbors, referred to as a $q$-lobby or $q$-panel, which are randomly selected from all neighbors of the given agent. The rationale behind this assumption is that, although individuals may have many social contacts, at any given moment they typically interact with only a small subset of them. If the $q$-lobby is unanimous, the focal agent adopts its shared opinion; otherwise, the agent’s update depends on the particular version of the model. Due to its analytical tractability, at least on complete and random graphs, and its conceptual clarity, the $q$-voter model has become one of the most widely used models of opinion dynamics\cite{Mobilia2015NonlinearZealots,Jedrzejewski2017PairNetworks,Vieira2018ThresholdModel,Vieira2020PairModel,Chmiel2020ANetworks,Gradowski2020PairNetworks,Krawiecki2024Q-voterApproximations,Weron2024CompositionOpinions}.
At this point, we would like to stress that, although the $q$-voter model is typically treated as a model of opinion dynamics, it can also be framed as a model of belief, attitude, or other related dynamics, since, as recently noted, the underlying processes are similar \cite{Olsson2024AnalogiesDynamics}.

In most studies, the $q$-voter model assumes symmetry between positive and negative opinions, meaning that both states are treated as equivalent. While this assumption is appropriate for modeling polarization or consensus, it is less realistic for innovation diffusion, where the two opinions, adoption and non-adoption, play inherently asymmetric roles. The first attempt to break this symmetry was made by Byrka et al. \cite{Byrka2016DifficultyPractices}. In their formulation, with a certain probability, an agent follows a unanimous group of $q$ neighbors, while with the complementary probability, it behaves independently, choosing opinions $+1$ or $-1$ with unequal probabilities: with some probability of engagement, the adoption state ($+1$) is chosen, and with the complementary probability, the non-adoption state ($-1$) is chosen. Their study, which focused on the diffusion of pro-environmental products and practices, assumed that the size of the $q$-lobby was $4$, and that the probability of engagement depended functionally on the perceived difficulty of the ecological behavior. Moreover, the model was analyzed analytically using only the mean-field approximation (MFA), and Monte Carlo simulations were performed only on complete and Watts-Strogatz networks \cite{Watts1998CollectiveNetworks}.

In the present work, we reformulate the model by treating the probability of engagement as an independent control parameter rather than assuming its functional dependence on behavioral difficulty. This modification increases the model’s universality, facilitates analytical treatment, and allows for direct comparison with the symmetric $q$-voter model with independence \cite{Nyczka2012PhaseDriving}. The main novelties of this study are as follows. First, we investigate the model for arbitrary values of $q$, a parameter known to play a crucial role in the dynamics of the symmetric version but previously fixed in the asymmetric case. Second, beyond the mean-field analysis, we extend the analytical approach by applying the pair approximation (PA) for the first time to an asymmetric $q$-voter model. Third, we complement the analytical results with Monte Carlo simulations on both Watts-Strogatz networks and empirical organizational networks to assess whether the PA provides reliable predictions in realistic topologies. To the best of our knowledge, the $q$-voter model has not previously been studied on organizational networks, nor has the validity of the PA been examined on such empirical structures.

The remainder of this paper is organized as follows. In the next section, we briefly review the related literature to place our study within a broader context. In particular, we describe all versions of the $q$-voter model with asymmetric opinions proposed to date and highlight the differences between these models and the one analyzed here. We also summarize previous applications of the PA in the context of the $q$-voter model, clearly illustrating the novelty of this work. In Section~\ref{sec:model}, we present the model in detail, then in Section~\ref{sec:methods} we develop the PA  for the asymmetric case. In Section~\ref{sec:results} we compare analytical results with Monte Carlo simulations on various network structures, including both artificial networks and empirical networks derived from organizational data. Finally, Section~\ref{sec:Conclusions} concludes the paper.  %In line with the call for methodological rigor, the NetLogo implementation of the model is available at \href{https://barbarakaminska.github.io/q-voter.model.github.io/}{https://barbarakaminska.github.io/q-voter.model.github.io/} and the C++ implementation at \href{https://github.com/Pitt14/Agent-Based-Innovation-Diffusion-Model}{https://github.com/Pitt14/Agent-Based-Innovation-Diffusion-Model}.

\section{Related works}
\label{sec:lit}
As we already mentioned, to the best of our knowledge, the first attempt to relax the assumption of symmetry between opinions in the $q$-voter model was proposed in the study on the diffusion of eco-innovation \cite{Byrka2016DifficultyPractices}. This paper was direct inspiration of the present work. However, since then, several other versions of asymmetric $q$-voter models have been introduced.

For example, the $q$-voter model under the influence of mass media has been studied on Barabasi-Albert and complete graphs using Monte Carlo simulations and the mean-field approximation \cite{Azhari2023TheGraph,Muslim2024MassModel,Fardela2024OpinionModel}. As in the original $q$-voter model, when the $q$-lobby is unanimous, the focal agent adopts the opinion shared by the $q$-lobby. If the $q$-lobby is not unanimous, the individual adopts the media’s opinion with probability $p$. Because in this formulation the focal agent always follows a unanimous $q$-panel, there is no room for independent innovativeness, which is essential for the diffusion of innovations \cite{Rogers2003DiffusionEdition}. Consequently, if all agents are initially in the non-adoption state, they remain non-adopters forever, unlike in the model with independence \cite{Byrka2016DifficultyPractices}.

The presence of media has also been studied by Civitarese \cite{Civitarese2021ExternalModels}. In this model, analyzed only on complete graphs, at each step an individual either follows the external field with probability $m$, representing the median opinion of society, or interacts locally with a randomly selected group of $q$ neighbors. An independence parameter $s$ captures media skepticism, defining the probability that an individual ignores the external field and instead flips their opinion randomly. When $s = 1$, meaning complete independence from media influence, the model reduces to the original $q$-voter model with independence \cite{Nyczka2012PhaseDriving}. Although this model incorporates independence, the independent behavior drives agents toward the median opinion rather than introducing novel opinions. Consequently, if all agents are initially in the non-adoption state, the median opinion is also non-adoption, and only minor fluctuations from this state may occur due to skepticism, so no true diffusion of innovation can be observed.

A different source of asymmetry was proposed by Mullick and Sen \cite{Mullick2025SocialInfluence}, who introduced the $q$-voter model with weighted influence. In this model, a randomly chosen agent adopts the unanimous opinion of a randomly selected $q$-panel. However, if the panel is not unanimous, the state of the focal agent depends nontrivially on the number of positive and negative agents, as individuals with positive opinions are assumed to have different influence weights than those with negative opinions. Another mechanism introducing asymmetry was proposed by Doniec et al. \cite{Doniec2025ModelingModel}. In their model, when the influence group is not unanimous, the composition of the group (i.e., the number of positive and negative agents) is irrelevant, and the probability that the focal agent changes its opinion depends solely on its current state. In neither of these two models is true independence present, since, in the case of a unanimous $q$-panel, the agent always adopts the opinion of that panel. 

As we can see, none of the models described above can adequately capture the diffusion of innovation, in the sense that if initially no one has adopted, then no one will ever adopt. Moreover, none of these models has been studied on empirical networks; in fact, most studies were conducted only on complete graphs. Finally, none of these models has been analyzed analytically within the pair approximation. On the other hand, several symmetrical versions of the $q$-voter model have been studied using PA, including: the $q$-voter model with independence \cite{Jedrzejewski2017PairNetworks}; the $q$-voter model with generalized anticonformity \cite{Abramiuk-Szurlej2021DiscontinuousGraphs}; the threshold $q$-voter model with independence, also known as the noisy threshold $q$-voter model \cite{Vieira2020PairModel}; the $q$-voter model with independence on multiplex networks \cite{Gradowski2020PairNetworks}; the $q$-voter model with quenched independence and quenched anticonformity \cite{Jedrzejewski2022PairNetworks}; the $q$-voter model with only conformity (without independence or anticonformity) for arbitrary values of $q$, including non-integer values, in both two-state and multi-state cases \cite{Ramirez2024OrderingModels}; and the $q$-voter model with independence on signed random graphs \cite{Krawiecki2024Q-voterApproximations}. Yet, in none of these cases has PA been validated on real-world networks.

\textbf{Therefore, in this paper we not only apply the pair approximation to an asymmetric $q$-voter model for the first time, but we also validate PA on empirical organizational networks for the first time, bridging the gap between analytical approaches and real-world social structures.}

\section{The model}
\label{sec:model}
We consider a population of $N$ agents, each occupying one node of an undirected network that represents the structure of social relations within a given social system, for example an organization. Every agent $x$ is characterized by a binary state variable $S_x = \pm 1$, where $S_x = +1$ denotes adoption (e.g., a positive opinion, acceptance or implementation of an innovation), and $S_x = -1$ represents non-adoption. The system evolves in discrete time steps according to two basic mechanisms: independent behavior and social influence, in the form of conformity.

\begin{figure*}[!ht]
     \centering
     \includegraphics[width=\textwidth]{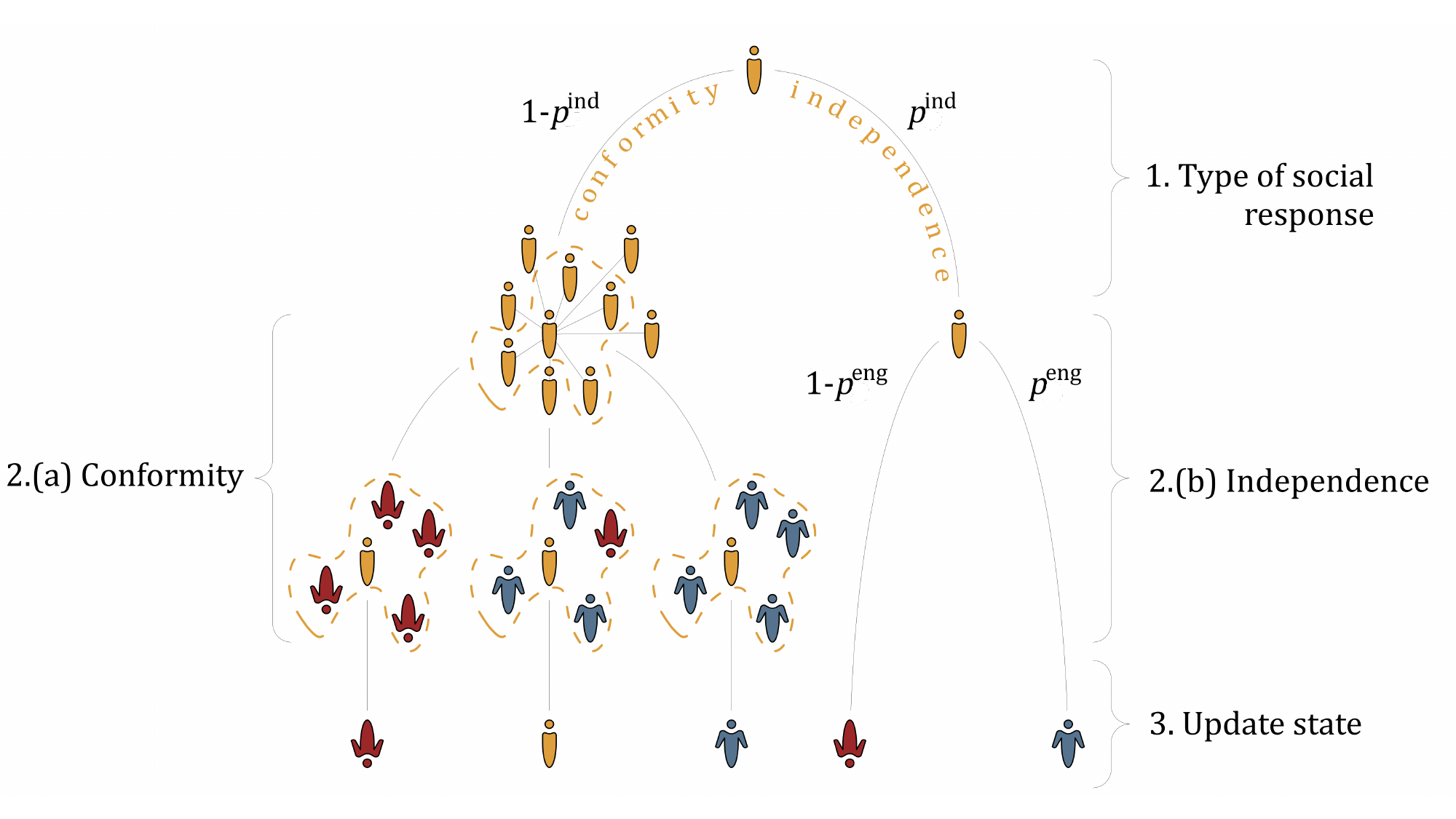}
     \caption{Illustration of the elementary update. In this specific example, the number of neighbors of agent $x$ is $9$, and the size of the influence group is $q=4$.}
     \label{fig:model}
\end{figure*}

At each time step, one randomly selected agent $x$ is chosen to potentially revise its state. With probability $p^{\mathrm{ind}}$, the agent acts independently of its neighbors. In this case, it adopts the innovation with probability $p^{\mathrm{eng}}$ or becomes non-adopted with probability $1 - p^{\mathrm{eng}}$. With complementary probability $1 - p^{\mathrm{ind}}$, the agent is subject to social influence. A group of $q$ neighbors is randomly selected without repetitions from its local neighborhood, that is, from agents directly connected to the focal agent. If all $q$ neighbors share the same opinion, the focal agent adopts that opinion; otherwise, it retains its current state. The parameter $q$ thus quantifies the size of the influence group and determines the strength of social pressure. The dynamics is given by the following update rule, illustrated in Fig. \ref{fig:model}:
\begin{enumerate}
    \item At a given time $t$, choose one agent $x$ at random.
    \item With probability $p^{\mathrm{ind}}$, the agent acts independently: it becomes adopter ($S_x = +1$) with probability $p^{\mathrm{eng}}$ or non-adopter ($S_x = -1$) with probability $1 - p^{\mathrm{eng}}$.
    \item Otherwise (with probability $1 - p^{\mathrm{ind}}$), select $q$ neighbors of agent $x$ at random without repetition (the $q$-lobby).
    \item If all $q$ neighbors share the same opinion ($q$-lobby is unanimous), the agent adopts that opinion; otherwise, it retains its current state.
    \item Time is updated as $t \to t + 1/N$.
\end{enumerate}

% \begin{figure*}[!ht]
%      \centering
%      \includegraphics[width=\textwidth]{SR_Article_Flowchart.pdf}
%      \caption{...}
%      \label{fig:flowchart}
%  \end{figure*} 

As mentioned already in the introduction, this model was inspired by the model of eco-innovation diffusion \cite{Byrka2016DifficultyPractices}, but it
can be also treated as the generalized $q$-voter model with independence, which  boils down to the original model for $p^{\mathrm{eng}}=1/2$ \cite{Nyczka2012PhaseDriving}. For $p^{\mathrm{eng}} \neq 1/2$ the model becomes asymmetric, which leads to irreversible hysteresis, as will be shown in Section \ref{sec:results}.

\section{Methods}
\label{sec:methods}
In this paper, we analyze the model using the following methods: Monte Carlo (MC) computer simulations and PA, which under additional assumptions can be reduced to MFA. We compare results obtained within all these methods both on artificial and empirical networks. 

\subsection{Monte Carlo simulations}
\label{ssec:sim}
We run simulations on both empirical and artificial networks, beginning with artificial networks generated using the Watts–Strogatz (WS) model, which is defined by three parameters \cite{Watts1998CollectiveNetworks}: 
%%%%%
\begin{enumerate}
    \item $N \in \mathbb{N}$: this parameter, which can take any natural number, denotes the size of a network (number of nodes) and is equal to the number of agents.    
    \item $\langle k \rangle \in \{0,\,\ldots,\,N-1\}$: this parameter, which takes an integer value between $0$ and $N-1$, is known as an average degree and it is related to a graph's density. Specifically, $\langle k \rangle$ is defined for undirected networks as the average number of neighbors of a node.  Empirical studies suggest that humans maintain around $150$ social relationships, organized in layers of decreasing emotional closeness, with average cumulative sizes of approximately $5$, $15$, $50$, and $150$ \cite{MacCarron2016CallingNumbers,Dunbar2024TheOn}.  Therefore, in contrast to \cite{Byrka2016DifficultyPractices}, we examine different values of $\langle k \rangle \in [5,\,150]$, inspired by empirical findings on social networks, and $\langle k \rangle = N-1$ to verify the model within the MFA.
    \item $\beta \in [0,\,1]$: this parameter, which takes a real value in the range $[0,\,1]$, represents the level of randomness in a network. It interpolates between a regular graph ($\beta = 0$) and a completely random graph ($\beta = 1$). Intermediate values of $\beta$ capture key structural features of real social networks, such as a low characteristic path length and a high clustering coefficient. These properties typically occur when $\beta \in (0.01,0.1)$ \cite{Watts1998CollectiveNetworks}.
\end{enumerate}
%%%%%
We chose the WS graph for two main reasons. First, as noted above, it captures key structural properties of real social networks within a certain parameter range. Second, it facilitates comparison with analytical results. When $\langle k \rangle = N-1$, the WS graph becomes a complete graph, for which MFA is exact. At $\beta = 1$, the graph resembles a random network with a low clustering coefficient, where the PA is expected to perform well. 

To conduct MC simulations on a WS graph, we first create an ensemble of $L$ graphs, each characterized by fixed values of the parameters $N$, $\langle k \rangle$ and $\beta$. For each graph $i = 1, \ldots, L$, we perform a total of $T$ Monte Carlo steps and calculate the final concentration of adopters, $c_i(T)$. We tested different values of $L$ and $T$, but present results primarily for the average over $L = 100$ samples and a stabilization time of $T = 2000$, since for these parameters the simulation results already agree well with the analytical predictions. After collecting results within a given ensemble, we calculate an ensemble average of the concentration of adopters:
\begin{equation}
     \langle c \rangle \equiv c = \frac{1}{L} \sum_{i=1}^L c_i(T).
 \end{equation}

\subsection{Analytical approach}
\label{ssec:pa}
%%%%%
We focus on two methods that allow us to reduce the ABM to the analytical model: MFA that has already been used to analyze similar model in \cite{Byrka2016DifficultyPractices}, and PA, which has already been applied to various binary state dynamics on complex networks \cite{GleesonBinary-StateBeyond}, in particular to different versions of the symmetrical $q$-voter model, as reviewed in Sec. \ref{sec:lit}.  PA is an enhanced version of the standard MFA. Therefore, here we will derive all formulas within the PA formalism and show that a special case of~the~results can be interpreted as a purely mean-field description of the model, which will be used here as a benchmark.

Within the PA, the time evolution of the model can be expressed by two differential equations describing the time evolution of the concentration of adopted agents $c$, i.e., agents in state $+1$, and of the concentration of active links (often called active bonds) $b$, i.e., links between two agents in~opposite states. As shown in \cite{Jedrzejewski2017PairNetworks,Jedrzejewski2019StatisticalSPOOF}, the general forms of these equations for an infinite system described by any binary state model, in particularly ours, are
%%%%
\begin{align}
\label{eq:general}
    \frac{d c}{d t} &= - \sum_{j \in \{-1, \, 1\}} c_j^{\vphantom{q}} \sum_k P(k) \sum_{i = 0}^k \binom{k}{i} \theta_{j}^i (1 - \theta_j^{\vphantom{q}})^{k - i} f(i,\,j,\,k), \nonumber \\
    \frac{d b}{d t} &= \frac{2}{\langle k \rangle} \sum_{j \in \{-1, \, 1\}} c_j^{\vphantom{q}} \sum_k P(k) \sum_{i = 0}^k \binom{k}{i} \theta_{j}^i (1 - \theta_j^{\vphantom{q}})^{k - i} f(i,\,j,\,k) (k - 2 i),
\end{align}
where $c_{j}$ is the concentration of agents in state $j \in \{-1$,~$1\}$ ($c_{1} = c \rightarrow c_{-1} = 1 - c$), $P(k)$ is the degree distribution of a considered network, $\theta_j \equiv P(-j|j)$ is the conditional probability of selecting an active link from all the links of the agent in state $j$,  i.e. the probability of choosing a neighbor in state $-j$, and $f(i,\,j,\,k)$ is a function describing the probability that an agent in state $j$ changes its state provided that $i$ out of its $k$ links are active.
%%%%

To calculate $\theta_j$ we use the definition of conditional probability
%%%
\begin{equation}
    \theta_j \equiv P(-j|j) = \frac{P(-j\,j)}{P(j)},
\end{equation}
where $P(-j\,j)$ is the probability of selecting first an agent in state $-j$ and then in state $j$, i.e. the probability of selecting an active link, and $P(j)$ is equal to the concentration $c_j$ of agents in state $j$. Hence, we can write the concentration of active links in the form
%%%%5
\begin{equation}
    b = \sum_{j \in \{-1,\,1\}} P(-j\,j).
\end{equation}
%%%%
There is the same number of connections between agents in states $(-1\,1)$ and $(1\,-1)$, because they share the same edge in the network. This implies that
%%%%
\begin{equation}
    P(-1\,1) = P(1\,-1) = \frac{b}{2}.
\end{equation}
%%%%
Having above results, we calculate conditional probabilities, as proposed in \cite{Jedrzejewski2017PairNetworks}
%%%%
\begin{equation}
    \theta_j = \frac{b}{2 c_j}.
\end{equation}
%%%%
In PA we assume that states of a chosen agent's neighbors are independent of each other. This assumption is equivalent to neglecting all closed (connected) triples or larger loops, or assuming that they can be approximated by pairs, hence the name of the method. Thus, the concentration $b$ of active links is the same for all nodes in the same state, no matter what is a degree of such a node, i.e., one value for all adopted agents, and the second value for all unadopted agents. Because we assume that neighbors of each agent are independent of each other, thus selecting $i$ active links follows a binomial distribution $\mathcal{B}(k,\,\theta_j)$ with the probability of success $\theta_j$ \cite{Jedrzejewski2017PairNetworks}.

In the case of highly clustered networks, the assumption of PA is certainly not fulfilled, because in such networks there are, by definition, many triples. In such a case, we expect PA to give incorrect results for ABM, which will be verified by comparing PA with the Monte Carlo simulations in Sec. \ref{sec:results}.

Within our model, the probability of changing a state due to independence, that occurs with probability $p^\mathrm{ind}$, is asymmetric and depends on the current state $j$ of the agent. We can write it in the form
%%%
\begin{equation}
    P(-j|j)_\text{ind} = \frac{(1+j)(1-p^\mathrm{eng}) + (1-j)p^\mathrm{eng}}{2}.
\end{equation}
%%%%
Note that substituting $j = -1$ in the above formula we obtain $P(1|-1)_\text{ind} = p^\mathrm{eng}$ and substituting $j = 1$ results in $P(-1|1)_\text{ind} = 1 - p^\mathrm{eng}$, which is consistent with the definition of the model. This part of the formula is the only new ingredient of the asymmetrical model with respect to the original $q$-voter model with independence. It does not depend on the state of~the~neighborhood and hence on $i$ or $k$ parameters, so it remains the same for MFA as well as PA approach.

The function $f(i$,~$j$,~$k)$ depends on the details of the particular model \cite{Jedrzejewski2019StatisticalSPOOF}. For the asymmetrical model considered in this paper $f(i$,~$j$,~$k)$ has not been calculated so far and thus will be derived below.

Changing state due to conformity occurs with probability $1-p^\mathrm{ind}$ if $q$ active links are selected. Here, there is a difference between MFA and PA. In the case of the pair approximation, the probability of selecting the first node in the opposite state is $i/k$, the probability of selecting the second one is $(i-1)/(k-1)$ and so on. Hence, the function takes the form
%%%%
\begin{align}
\label{eq:f:pa}
    f(i,\,j,\,k) &= (1-p^\mathrm{ind})\frac{i! (k-q)!}{k! (i-q)!} + p^\mathrm{ind} P(-j|j)_\text{ind}\nonumber\\
    &= (1-p^\mathrm{ind})\frac{i! (k-q)!}{k! (i-q)!} + p^\mathrm{ind} \frac{1 + j (1 - 2 p^\mathrm{eng})}{2}.
\end{align}
%%%
In the case of the mean-field approximation, it is assumed that the probability of selecting a node in the opposite state equals $c_{-j}^{\vphantom{q}}$. Hence, the function takes the form
%%%%
\begin{align}
\label{eq:f:mfa}
    f(i,\,j,\,k) &= (1-p^\mathrm{ind}) c_{-j}^q + p^\mathrm{ind} P(-j|j)_\text{ind} \nonumber\\
    &= (1-p^\mathrm{ind}) c_{-j}^q + p^\mathrm{ind} \frac{1 + j (1 - 2 p^\mathrm{eng})}{2}.
\end{align}
%%%%
Summing over $i$ and $k$ in Eqs. \eqref{eq:general} we obtain for the pair approximation
%%%%
\begin{align}
\label{eq:dcdb:pa}
    \frac{dc}{dt} &= - \sum_{j} c_{j} \left[(1-p^\mathrm{ind}) \theta_{j}^q + p^\mathrm{ind} \frac{1 + j (1 - 2 p^\mathrm{eng})}{2}\right] j, \nonumber\\
    \frac{db}{dt} &= \frac{2}{\langle k \rangle} \sum_{j} c_j^{\vphantom{q}} \Bigl[{\vphantom{\frac{p^\mathrm{ind}}{2}}}(1-p^\mathrm{ind}) \theta_j^q [\langle k \rangle - 2q - 2(\langle k \rangle - q) \theta_j^{\vphantom{q}}] \\
    &+ p^\mathrm{ind} \langle k \rangle \frac{1 + j(1 - 2 p^\mathrm{eng})}{2} (1 - 2 \theta_j^{\vphantom{q}})\Bigr], \nonumber
\end{align}
%%%%%
and for the mean-field approximation
%%%%%
\begin{align}
\label{eq:dcdb:mfa}
    \frac{dc}{dt} &= - \sum_{j} c_{j}^{\vphantom{q}} \left[(1-p^\mathrm{ind}) c_{-j}^q + p^\mathrm{ind} \frac{1 + j (1 - 2 p^\mathrm{eng})}{2}\right] j,\nonumber\\
    \frac{db}{dt} &= 2 \sum_{j} c_{j}^{\vphantom{q}} \left[(1-p^\mathrm{ind}) c_{-j}^q + p^\mathrm{ind} \frac{1 + j (1 - 2 p^\mathrm{eng})}{2}\right] (1 - 2 \theta_{j}^{\vphantom{q}}).
\end{align}

It is worth noting that the differential equation describing the concentration of adopted agents $c$ within the MFA approach does not depend on the concentration of active links $b$, see Eqs. (\ref{eq:dcdb:mfa}). Therefore, in this case, the equations for $c$ and $b$ can be solved independently. Note also that Eqs. (\ref{eq:dcdb:pa}) for PA depend only on the average degree of a node $\langle k \rangle$, not on the entire degree distribution of the network $P(k)$, analogously to \cite{Jedrzejewski2017PairNetworks}. This follows from the specific form that the function $f(i, j, k)$ takes for the $q$-voter model, which, after summation over $i$, leads to a formula linear in $k$. Recently, it has been shown that such a simple result is obtained only for the $q$-voter model in which $q$ neighbors are chosen without repetitions. If repetitions are allowed, then the result depends on $P(k)$ \cite{Ramirez2024OrderingModels}.

Eqs. (\ref{eq:dcdb:pa}) and (\ref{eq:dcdb:mfa}) allow to obtain the time evolution of the system, but also the stationary states, which are determined by the conditions
%%%%
\begin{equation}
\label{eq:stationary}
    \frac{dc}{dt} = 0 \quad \wedge \quad \frac{db}{dt} = 0.
\end{equation}
%%%%%
It is possible to derive the explicit formulas for $p^\mathrm{ind}(c)$ and $b(c)$ within MFA and PA.  Additionally, for MFA, we can also obtain the analytical formula for $p^\mathrm{eng}(c)$, whereas within PA the solution is implicit. For the pair approximation 
%%%%%
\begin{align}
\label{eq:all:pa}
    \vphantom{\Bigl[} p^\mathrm{ind} &=  \frac{c \theta_{1}^q - (1-c)\theta_{-1}^q}{c \theta_{1}^q - (1-c) \theta_{-1}^q + p^\mathrm{eng} - c}, \nonumber\\
    \vphantom{\Bigl[} b &= 2 \frac{c (1-c)[(1-c)^q p^\mathrm{eng} - c^q (1-p^\mathrm{eng})] - \frac{q}{\langle k \rangle}(p^\mathrm{eng}-c)[c (1-c)^q + (1-c) c^q]}{(1-c)^q p^\mathrm{eng} - c^q (1-p^\mathrm{eng}) - \frac{q}{\langle k \rangle} (p^\mathrm{eng}-c) [(1-c)^q + c^q]},
\end{align}
%%%%%
and for the mean-field approximation
%%%%%
\begin{align}
\label{eq:all:mfa}
    \vphantom{\Bigl[} p^\mathrm{ind} &= \frac{c(1-c)^q - (1-c)c^q}{c(1-c)^q - (1-c)c^q + p^\mathrm{eng} - c}, \nonumber\\
    \vphantom{\Bigl[} p^\mathrm{eng} &= \frac{(1-p^\mathrm{ind})[c(1-c)^q - (1-c)c^q] + p^\mathrm{ind} c}{p^\mathrm{ind}}, \\
    \vphantom{\Bigl[} b &= 2 c (1 - c).\nonumber
\end{align}
%%%%%
If we put $p^\mathrm{eng} = 1/2$ into Eqs. \eqref{eq:all:pa} we obtain the same formula for $p^\mathrm{ind}(c)$ as for the model with symmetric independence \cite{Jedrzejewski2017PairNetworks}. Note also that taking $\langle k \rangle \to \infty$ we obtain an infinite complete graph, which corresponds to the MFA approach.

%%%%
\section{Results}
\begin{figure*}[!ht]
    \centering
    \includegraphics[width=1\textwidth]{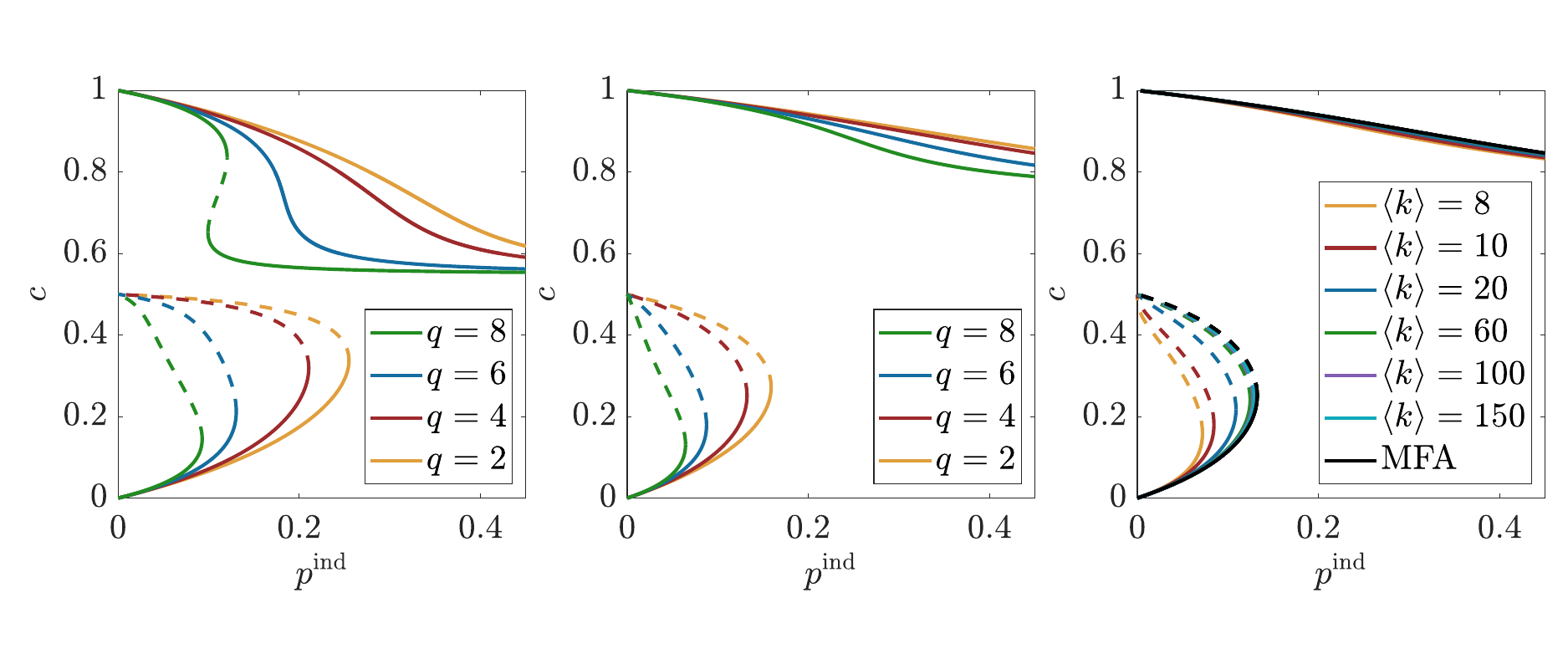}
    \caption{The stationary concentration of adopters $c$ as a function of the probability of independence $p^{\mathrm{ind}}$, obtained within MFA and PA for different values of the probability of engagement $p^{\mathrm{eng}}$, the influence group size $q$, and the network degree $\langle k \rangle$. 
    \textbf{Left panel:} MFA results for $p^{\mathrm{eng}} = 0.55$ and the values of $q$ indicated in the legend. \textbf{Middle panel:} Same as in the left panel, but for $p^{\mathrm{eng}} = 0.75$. \textbf{Right panel:} PA results for the stationary concentration of adopters $c$ as a function of $p^{\mathrm{ind}}$ for $p^{\mathrm{eng}} = 0.75$, $q = 4$, and different values of $\langle k \rangle$ indicated in the legend.}
    \label{fig:Res:analit}
\end{figure*}

\begin{figure*}[!ht]
    \centering
    \includegraphics[width=1\textwidth]{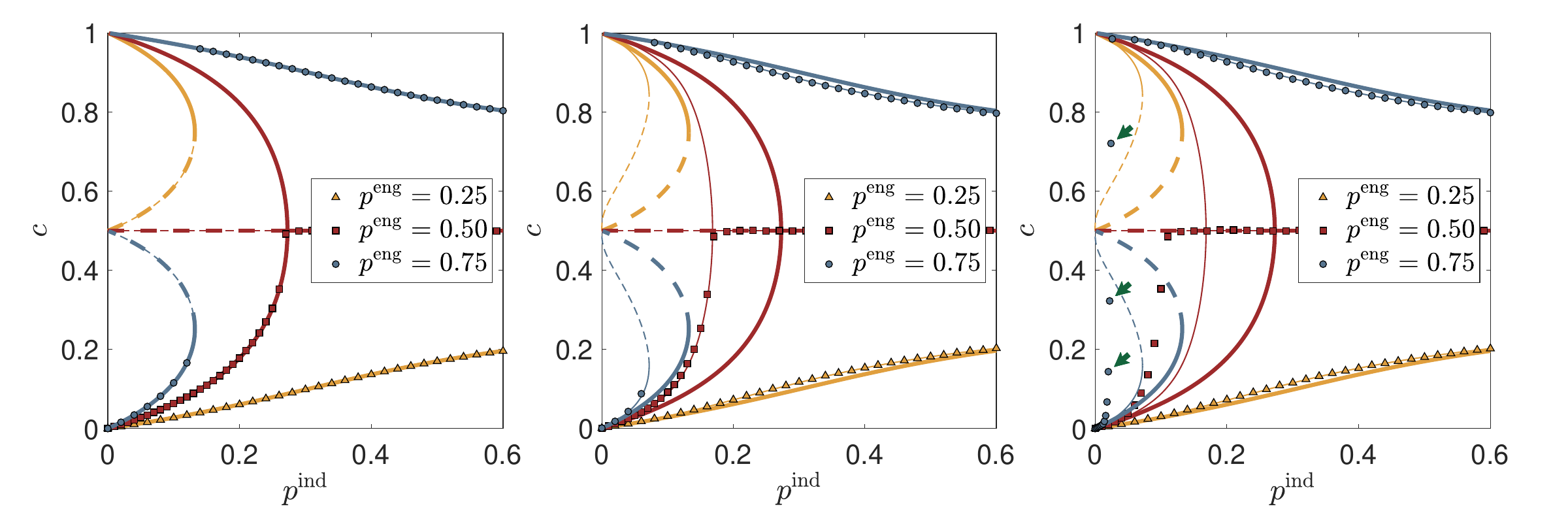}
    \caption{Stationary concentration of adopters $c$ as a function of the independence probability $p^\mathrm{ind}$ for the influence group size $q = 4$ and several values of the engagement probability $p^\mathrm{eng}$ (indicated in the legends). Each panel corresponds to a different network structure. \textbf{Left panel:} complete graph. \textbf{Middle panel:} Watts–Strogatz (WS) graph with average degree $\langle k \rangle = 8$ and rewiring probability $\beta = 1$. \textbf{Right panel:} WS graph with $\langle k \rangle = 8$ and $\beta = 0.05$. Thin and thick lines represent analytical results from PA and MFA, respectively, while symbols denote the outcomes of MC simulations starting from an initial state with all agents unadopted ($c = 0$), for a system of size $N = 10^4$. In case of a complete graph and WS graph with $\beta=1$ stabilization time $T = 2000$ (used in the left and middle panels) is sufficient to reach the stationary state and MC results overlap analytical ones. However, in case of WS graph with $\beta=0.05$ even a twice longer time $T = 4000$ (used in the right panel) is not sufficient for small values of $p^\mathrm{ind}$, as the system has not reached yet the stationary state. All points denoted by arrows will eventually reach the stationary state, which lies on the upper brunch of the analytical solution.} 
    \label{fig:Cp_diff_k_PA_MCS_arrows}
\end{figure*}

Analytical formulas derived in the previous sections can be used to obtain the relation between the stationary concentration of adopters $c$ and the other parameters of the models. For example, from Eq.~\eqref{eq:all:mfa} we can obtain $p^\mathrm{ind}=p^\mathrm{ind}(c)$, and by inverting this relation we can obtain $c=c(p^\mathrm{ind})$ within MFA, as shown in the left and middle panels of Fig.~\ref{fig:Res:analit}. To obtain results from PA, presented in the right panel of Fig.~\ref{fig:Res:analit}, we used Eq.~\eqref{eq:all:pa}. For $p^\mathrm{eng}=0.5$, our model reduces to the $q$-voter model with independence, and the relation $c=c(p^\mathrm{ind})$ is symmetrical around $c=1/2$, which can be seen in Fig. \ref{fig:Cp_diff_k_PA_MCS_arrows}. There is a continuous phase transition between the state with a majority ($c \ne 0$) and the state without a majority for $q<5$, which becomes discontinuous for $q>5$, with a tricritical point at $q=5$~\cite{Nyczka2012PhaseDriving}.

However, if we break the symmetry between states $+1$ and $-1$, i.e., set $p^\mathrm{eng} \neq 0.5$, the transition becomes discontinuous even for small values of $q>1$, as shown in the left panel of Fig.~\ref{fig:Res:analit} for $p^\mathrm{eng}=0.55$. In such a case, we obtain several qualitatively different stationary solutions. Let us first focus on $q=2$, shown in the left and middle panels of Fig.~\ref{fig:Res:analit}. There are three possible stationary solutions: two stable ones, indicated by solid lines, and one unstable, indicated by the dashed line. The unstable state can be treated as a critical mass. Below this concentration of adopters, the system eventually reaches the final state with a majority of non-adopters, while above it, the system reaches the final state with a majority of adopters. However, for the probability of engagement $p^\mathrm{eng}=0.55$ (left panel in Fig.~\ref{fig:Res:analit}), it is difficult to speak about the diffusion of innovation, in the sense that the final concentration of adopters is high only for very small values of $p^\mathrm{ind}$, and at the same time, the critical mass is very large. For $p^\mathrm{eng}=0.75$, the results resemble much more what can be expected for the diffusion of innovation. In the right panel of Fig.~\ref{fig:Res:analit}, we see that when we take into account influence only from the closest social circles, $\langle k \rangle = 8$ and $\langle k \rangle = 10$, the critical mass becomes even smaller. Interestingly, in all these cases we observe irreversible hysteresis: once the concentration of adopters becomes larger than the critical mass, the adoption steady state is reached and cannot be left, even if the level of independence is decreased.  

\label{sec:results}

\begin{figure*}[!ht]
    \centering
    \includegraphics[width=1\textwidth]{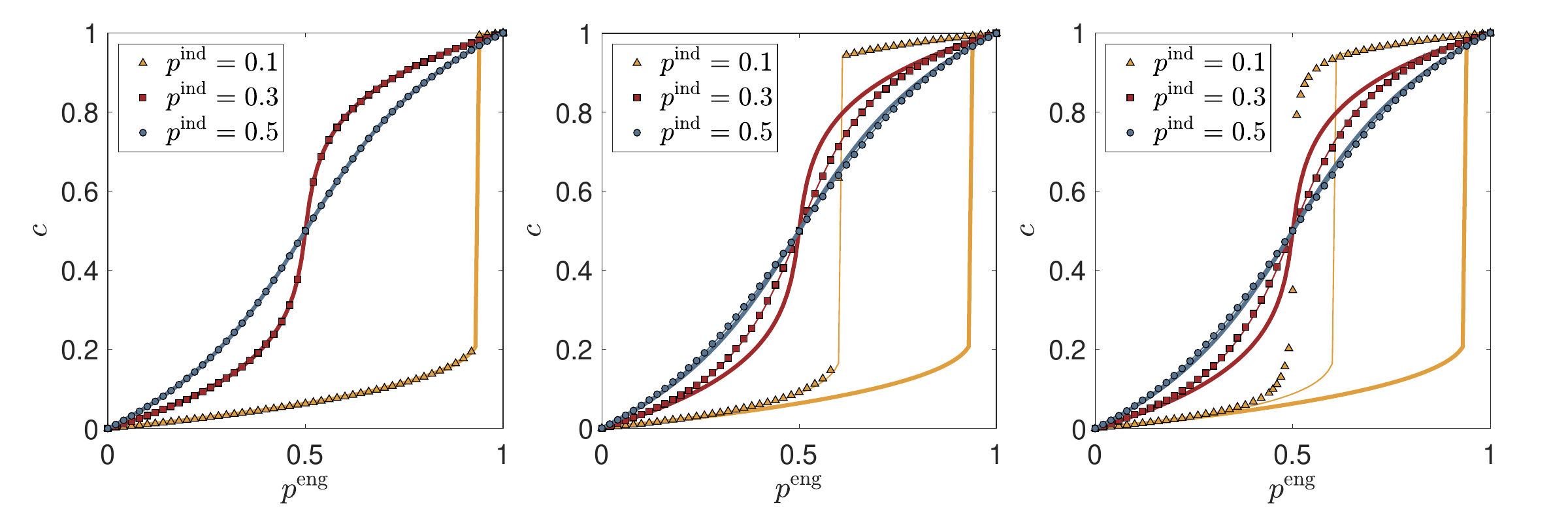}
    \caption{Stationary concentration of adopters $c$ as a function of the engagement probability $p^\mathrm{eng}$ for the influence group size $q = 4$ and several values of the independence probability $p^\mathrm{ind}$ (indicated in the legends). Each panel corresponds to a different network structure, analogously as in Fig. \ref{fig:Cp_diff_k_PA_MCS_arrows}. \textbf{Left panel:} complete graph. \textbf{Middle panel:} (middle) WS graph with average degree $\langle k \rangle = 8$ and rewiring probability $\beta = 1$. \textbf{Right panel:} WS graph with $\langle k \rangle = 8$ and $\beta = 0.05$. Thin and thick lines represent analytical results from PA and MFA, respectively, while symbols denote the outcomes of Monte Carlo simulations starting from an initial state with all agents unadopted ($c = 0$), for a system of size $N = 10^4$ and stabilization time $T = 2000$.} 
    \label{fig:Cf_diff_p_PA_MCS}
\end{figure*}

Now let us compare the analytical results with Monte Carlo simulations on complete and WS graphs. As expected, for the complete graph (left panels in Figs.~\ref{fig:Cp_diff_k_PA_MCS_arrows} and \ref{fig:Cf_diff_p_PA_MCS}), PA and MFA overlap, and the simulation results agree very well with the analytical predictions. However, for the WS graph with $\beta = 1$, PA significantly improves upon MFA, at least for sparse networks, as shown in the middle panels of Figs.~\ref{fig:Cp_diff_k_PA_MCS_arrows} and \ref{fig:Cf_diff_p_PA_MCS}. As expected, for small-world networks ($\beta = 0.05$), PA still performs better than MFA but does not reproduce the simulation results as accurately as for larger values of $\beta$, as shown in the right panels of Figs.~\ref{fig:Cp_diff_k_PA_MCS_arrows} and \ref{fig:Cf_diff_p_PA_MCS}. These are all expected results and agree well with the PA results for the symmetrical versions of the $q$-voter model with independence \cite{Jedrzejewski2017PairNetworks}. The question that has never been asked before in the context of the $q$-voter model is how well PA agrees with the simulation results on real social networks.

Before answering this question, we would like to highlight a particularly interesting result from the perspective of the diffusion of innovation, which is shown in Fig.~\ref{fig:Cf_diff_p_PA_MCS}. First, we observe that even a very small level of independence, $p^{\mathrm{ind}} = 0.1$, which can be related to the level of innovativeness in Rogers’ theory of diffusion of innovations (i.e., individuals who adopt without social influence), can lead to nearly full adoption, provided that the level of engagement is high enough. What is even more interesting is that the required level of engagement must be much higher on a complete graph than on networks with lower density. This suggests that the interplay between individual innovativeness and network structure is crucial for understanding how new behaviors or ideas spread.

\subsection{Results on organizational networks}
\begin{figure*}[!ht]
    \centering
    \includegraphics[width=\textwidth]{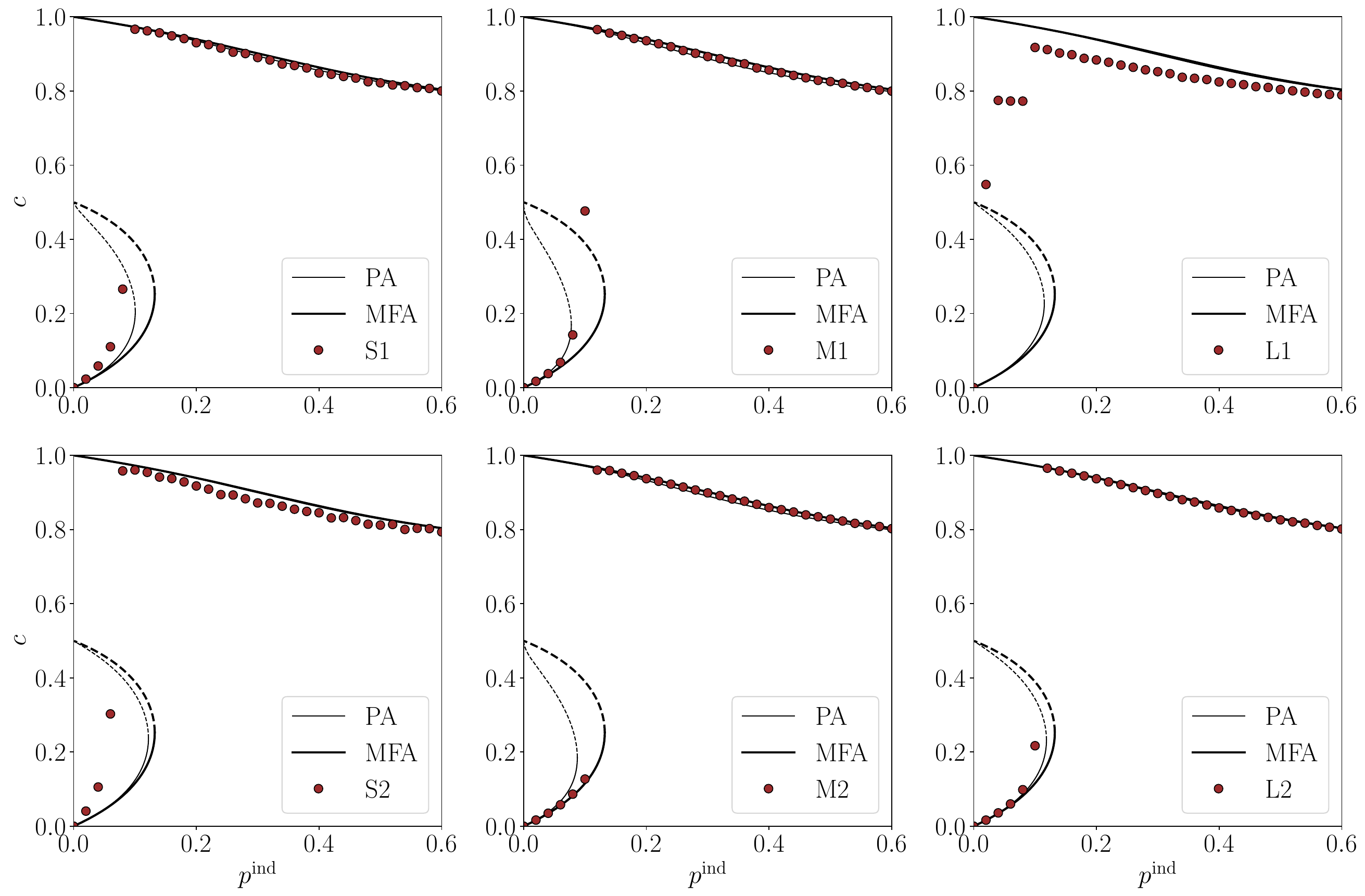}
    \caption{Stationary concentration of adopted $c$ as a function of the probability of independence $p^\mathrm{ind}$ for the size of the influence group $q = 4$ and the engagement probability $p^\mathrm{eng} = 0.75$. Each panel corresponds to a different empirical social network, as indicated in the legends. Thin and thick lines represent analytical results from PA and MFA, respectively, while symbols denote results from Monte Carlo simulations for the empirical social networks with stabilization time $T = 2000$. Simulation results are averaged over $100$ samples.} 
    \label{fig:Stationary_p_eng_0.75_q_4_min_q_k}
\end{figure*}

Now we answer the question how properly MFA and PA describes the simulation results on real networks. Therefore, we run the model on six real organizational networks: two representing small-sized (S) companies, two representing medium-sized (M) companies, and two representing large-sized (L) companies, and compare simulation results with analytical approximations. The organizational datasets were collected by Fire \textit{et al.} from employees' Facebook pages~\cite{Fire2016OrganizationNetworks} and made publicly available to researchers at \href{https://data4goodlab.github.io/dataset.html}{https://data4goodlab.github.io/dataset.html}. Because in real networks it may happen that a given node $x$ has a degree $k_x < q$, we slightly reformulate the algorithm: instead of taking $q$ neighbors, we take $\min(q, k_x)$ neighbors. The results are shown in Fig.~\ref{fig:Stationary_p_eng_0.75_q_4_min_q_k}. The PA still performs quite well, although its accuracy depends on the specific network. 

Our first step in understanding why PA performs better or worse on a given network was to compute the typical characteristics of these networks. In Tab.~\ref{tab:Networks_parameters}, we present some of these characteristics, including the average degree $\langle k \rangle$ and its standard deviation $\sigma_{\langle k \rangle}$, the average shortest path length $\langle L \rangle$ and its standard deviation $\sigma_{\langle L \rangle}$, the average of local clustering coefficients (average local clustering) $\langle C \rangle$ and its standard deviation $\sigma_{\langle C \rangle}$, as well as the global clustering coefficient (global clustering) $C_g$. We calculated both clustering coefficients, $ \langle C \rangle $ and $ C_g $, because they are often treated interchangeably despite capturing different aspects of network structure \cite{Prokhorenkova2014GlobalNetworks,Estrada2016WhenDiverge}. The global clustering coefficient measures the overall tendency of nodes to form interconnected groups in the network. It is based on counting "closed triplets", which are sets of three nodes where each node is connected to the other two, forming a triangle. The ratio of these closed triplets to all possible connected triplets in the network determines the global clustering coefficient. In contrast, the average local clustering coefficient quantifies how well a node's neighbors are interconnected by comparing the number of actual links between them to the maximum possible. It was shown that for Watts-Strogatz (WS) graphs, $C_g$ is very close to $ \langle C \rangle $ \cite{Barrat2000OnModels}. However, this relationship does not hold for arbitrary networks, where significant differences between the two measures can arise \cite{Prokhorenkova2014GlobalNetworks,Estrada2016WhenDiverge}. In networks with a heterogeneous degree distribution, the global clustering coefficient $C_g$ is often much lower than the average local clustering coefficient $ \langle C \rangle $ because high-degree nodes tend to exhibit lower clustering. We wanted to check whether this phenomenon is also present in the empirical networks we analyzed.  

Furthermore, to gain deeper insight into the possible differences between the results obtained for WS graphs and empirical networks, we have included in the last column of Tab.~\ref{tab:Networks_parameters} the rewiring probability $\beta$ corresponding to a WS graph with the same clustering coefficient $\widetilde{C}(\beta)$ and average degree $\langle k \rangle$ as the analyzed network. This value was calculated using the formula provided in \cite{Barrat2000OnModels}: 

%\textcolor{red}{approximating the average local clustering coefficient for WS graph by}: 
%%%%%
\begin{equation}
\label{eq:C_beta}
    C(\beta) \sim \widetilde{C}(\beta) = C(0)(1 - \beta)^3 = \frac{3(\langle k \rangle - 1)}{2(2 \langle k \rangle - 1)}(1 - \beta)^3,
\end{equation}
%%%%%
where $\widetilde{C}(\beta)$ is defined as the ratio of the mean number of links between the neighbors of a given node to the mean number of possible links between them. Hence,
\begin{equation}
    \label{eq:beta}
    \beta = 1 - \sqrt[3]{\frac{\widetilde{C}(\beta)}{C(0)}}.
\end{equation}
%%%%%%
\begin{table}[!ht]
    \centering
    \caption{Quantities of empirical social networks from organizations: the number of agents $N$, the average degree $\langle k \rangle$ with the standard deviation $\sigma_{\langle k \rangle}$, the average shortest path length $\langle L \rangle$ with the standard deviation $\sigma_{\langle L \rangle}$, 
    the average clustering coefficient $\langle C \rangle$ with the standard deviation $\sigma_{\langle C \rangle}$,  the global clustering coefficient $C_g$, and the rewiring probability $\beta$ corresponding to a WS graph with the same values of parameters as the network.}
    \label{tab:Networks_parameters}
    \begin{tabular}{c*{10}{r}}
        \\\toprule
        & \multicolumn{10}{c}{Global parameter} \\
        \cmidrule(lr){2-11}
        Network & \multicolumn{1}{c}{$N$} & \multicolumn{1}{c}{$\langle k \rangle$} & \multicolumn{1}{c}{$\sigma_{\langle k \rangle}$} & \multicolumn{1}{c}{$\langle L \rangle$} & \multicolumn{1}{c}{$\sigma_{\langle L \rangle}$} & \multicolumn{1}{c}{$\langle C \rangle$} & \multicolumn{1}{c}{$\sigma_{\langle C \rangle}$} & \multicolumn{1}{c}{$C_g$} & \multicolumn{1}{c}{$\widetilde{C}(\beta)$} & \multicolumn{1}{c}{$\beta$}\\
        \midrule
        S1 & $320$ & $14.81$ & $14.26$ & $2.70$ & $0.82$ & $0.49$ & $0.25$ & $0.29$ & $0.29$ & $0.265$ \\
        S2 & $165$ & $8.80$ & $8.33$ & $2.86$ & $0.96$ & $0.43$ & $0.30$ & $0.33$ & $0.33$ & $0.224$ \\
        M1 & $1429$ & $27.09$ & $28.72$ & $3.03$ & $0.85$ & $0.42$ & $0.19$ & $0.26$ & $0.26$ & $0.292$ \\
        M2 & $3862$ & $45.22$ & $29.58$ & $2.76$ & $0.53$ & $0.31$ & $0.13$ & $0.23$ & $0.23$ & $0.328$ \\
        L1 & $5793$ & $10.62$ & $18.39$ & $6.26$ & $2.05$ & $0.17$ & $0.26$ & $0.26$ & $0.26$ & $0.281$ \\
        L2 & $5524$ & $34.11$ & $31.81$ & $3.50$ & $1.03$ & $0.36$ & $0.19$ & $0.22$ & $0.22$ & $0.330$ \\
        \bottomrule
    \end{tabular}
\end{table}

We see in Tab. \ref{tab:Networks_parameters} that, indeed, for most of the networks we analyzed, $C_g < \langle C \rangle$, but not for L1, where $C_g > \langle C \rangle$. From Tab. \ref{tab:Networks_parameters} we can notice also that L1 has a relatively low average degree of $10.62$ with high standard deviation, indicating that some nodes have significantly more connections than others. Additionally, L1 has an unusually long shortest path, further suggesting a hierarchical topology. These structural characteristics contribute to an inflated global clustering coefficient. Stars themselves have low local clustering (because the nodes around the hub are not connected), but if the hubs are connected to each other, they can form many global triangles. $C_g$ considers all triads in the network. If most of the open triads are closed (e.g. because the hubs are connected to each other), then $C_g$ can be large. On the other hand, $\langle C \rangle$ calculates the average local clustering for each node separately. If most of the nodes are leaves in the stars (i.e. have zero clustering), then $\langle C \rangle$ can be low.

The fact that we observe the unusual relationship $C_g > \langle C \rangle$ only for L1 is particularly interesting when we examine Fig.~\ref{fig:Stationary_p_eng_0.75_q_4_min_q_k}, where the simulation results in the empirical networks are compared with the analytical predictions. For this network, the agreement between simulations and analytical results is the weakest. Initially, we expected the poorest agreement for the smallest network, S2, which consists of only $N = 165$ nodes. Indeed, for this network, the simulation and analytical results also diverge, though to a lesser extent than for L1, in the sense that a critical point can still be identified above which both results converge. In contrast, for the L1 network, no clear criticality is observed and adoption (even if only partial) occurs for arbitrarily small values of $p^\mathrm{ind}$.

Then we thought that there is another, much simpler explanation why the PA does not produce accurate results for L1. The PA result in Fig.~\ref{fig:Stationary_p_eng_0.75_q_4_min_q_k} corresponds to a lobby of size $q=4$. However, the effective lobby size, calculated as
\begin{equation}
q_\text{eff} = \frac{1}{N} \sum_{i=1}^{N} \min(q,k_i),
\end{equation}
takes the following values for the considered networks:
S1: $q_\text{eff} = 3.809375$, S2: $q_\text{eff} = 3.339394$, M1: $q_\text{eff} = 3.966410$, M2: $q_\text{eff} = 4.000000$, L1: $q_\text{eff} = 2.611255$, L2: $q_\text{eff} = 3.999638$. As can be seen, for most networks $q_\text{eff}$ is close to $4$, while for L1 it is substantially lower, which could provide a simple explanation for why PA fails in this case.  

However, when we consider the influence of $q$ on the PA results, shown in Fig.~\ref{fig:Res:analit}, we see that for $q=2$ the critical point $p_c^{\text{ind}}(q)$ above which innovation can spread is actually higher than for $q=4$. This is the opposite of what is observed for L1 in Fig.~\ref{fig:Stationary_p_eng_0.75_q_4_min_q_k}, where the critical point from simulations appears to be lower than that predicted by PA. Therefore, this simple explanation is not satisfactory. A much better, yet still simple, explanation is that due to the specific structure of L1, there are many nodes of degree $1$, and for $q=1$ there is no phase transition, and the system always reaching the adopted state. This indeed agrees with the result obtained for L1.

\section{Conclusions}
\label{sec:Conclusions}
Real-world diffusion of innovation is often unpredictable: some promising ideas spread rapidly and irreversibly through societies or organizations, while others, despite clear advantages, fail to gain traction or disappear after initial success. For example, the diffusion of hybrid corn among American farmers and the rapid global adoption of mobile phones illustrate successful innovations that reached a critical mass and became irreversible. In contrast, many technically sound ideas, such as the Dvorak keyboard, failed to diffuse widely despite clear functional advantages \cite{Rogers2003DiffusionEdition}. 

Our study provides a theoretical description of these phenomena within the biased-independence $q$-voter model, a generalization of the basic $q$-voter model with independence \cite{Nyczka2012PhaseDriving}. In this model, agents adopt innovations under the competing influences of conformity and independent choices, with an engagement parameter introducing asymmetry between adoption and non-adoption. This simple mechanism reproduces key empirical features of innovation diffusion: discontinuous phase transitions, critical mass effects, and irreversible hysteresis. Specifically, we find that once the concentration of adopters exceeds a certain critical mass, the system collectively adopts the innovation, although full ($c=1$) adoption is not achieved. Moreover, irreversible hysteresis implies that once a population becomes predominantly adopted, it does not return to the non-adopted state even when the level of independence, interpretable as the degree of innovativeness or openness to change, is reduced.

We acknowledge that this is a very simple model and only one of many possible approaches to describing innovation diffusion \cite{Rand2021AgentbasedCriticisms}; therefore, it should not be treated as the main contribution of this work. However, from a modeling perspective, the results extend our theoretical understanding of opinion dynamics. In the symmetric $q$-voter model with independence, discontinuous transitions occur only for large influence groups ($q>5$), whereas in the present asymmetric formulation, they appear for all values of $q>1$. It is an interesting and previously overlooked result, given that in many real situations some level of asymmetry occurs. On the other hand, hysteresis, which accompanies discontinuous phase transitions, appears in many psycho-social systems \cite{Sznajd-Weron2024TowardModeling}. Our results suggest that one possible reason for this is a bias toward one of the two choices.

Methodologically, this study provides two key advances. First, we developed the pair approximation (PA) for an asymmetric $q$-voter model for the first time. Second, we validated the PA and mean-field approximation (MFA) on empirical organizational networks, showing that the accuracy of a given analytical method depends on the particular network, and in some cases PA can give surprisingly accurate results, even on small networks. This last finding was particularly striking, because until now we believed that PA could give reliable results only on random graphs.  

Our research, of course, has several limitations. First, we considered only one specific model, and we have already observed that the PA performs much worse for the $q$-voter model with anticonformity than for the model with independence \cite{Abramiuk-Szurlej2021DiscontinuousGraphs,Jedrzejewski2022PairNetworks}. In future work, it would be interesting to validate the PA for other opinion dynamics models on the same set of empirical networks to determine whether the accuracy of the PA depends more on the network structure or on the details of the model. Second, we considered only one specific set of empirical networks. Nevertheless, we hope that our paper will serve as a call to validate analytical methods on empirical networks, rather than solely on theoretical graphs, since it may turn out that, perhaps surprisingly, the PA yields unexpectedly accurate results in real-world settings.

%\bibliography{references}

\section*{Funding}
K.S.-W. was supported by the National Science Center (NCN, Poland) under grant no. 2019/35/B/HS6/02530, and A.A.-S. was supported by the Polish Ministry of Science and Higher Education through the “Diamentowy Grant” project no. DI2019 015049.

\section*{Acknowledgments}
A.A.-S. would like to thank Patryk Siedlecki for his assistance in developing and optimizing the C++ implementation of the model.

%%%%%
\section*{Declaration of generative AI and AI-assisted technologies in the writing process}
%%%%
During the preparation of this work the authors used DeepL Write and ChatGPT in order to improve the readability and language of certain parts of the work. After using these tools, the authors reviewed and edited the content as needed and take full responsibility for the content of the publication.

\section*{Author contributions statement}
A.A.-S. performed all analytical calculations, analyzed the networks, and ran and analyzed the computer simulations; K.S.-W. designed and supervised the work and wrote the manuscript.

\section*{Competing interests}
The authors declare no competing interests.

\end{document}